\documentstyle[prl,aps,preprint,epsf]{revtex} 
\large  

\begin{document}
\draft

\title{Three manifestations of the pulsed harmonic potential} 
\author{T. J. Newman and R. K. P. Zia} 
\address{Department of Physics,\\ 
Virginia Polytechnic Institute and State University,\\ 
Blacksburg VA 24061,\\ USA} 
\maketitle
\begin{abstract}
We consider, in turn, three systems being acted upon by a 
regularly pulsed harmonic potential (PHP). These are i)
a classical particle, ii) a quantum particle, and iii)
a directed line. We contrast the mechanics of the
first two systems by parameterizing their bands of stability and
periodicity. Interesting differences due to quantum fluctuations
are examined in detail. The fluctuations of the directed line
are calculated in the two cases of a binding PHP, and an
unbinding PHP. In the latter case there is a finite maximum
line length for a given potential strength.

\end{abstract}
\vspace{5mm} \pacs{PACS numbers: 03.20.+i, 03.65.-w, 05.40.+j, 74.40.+k}

\newpage

\section{Introduction}

Within the modelization of many problems in mechanics, the 
potential is taken to be time-independent, and sets the stage
for the ensuing particle dynamics. Quite generally, however, the 
potential may have its own (parametric) dynamics: the stage is
shifting beneath the actors' feet, so to speak. In quasi-static
problems this parametric dynamics will be on such slow time scales
that it may be safely ignored; or else, an adiabatic treatment
may be used\cite{ditt}. On the other hand, if the time scale
is of exceedingly rapid character, it may be easier to forget
the potential altogether, and model the parametric variation
through some fast degrees of freedom. The Langevin equation
represents a possible outcome of such a procedure\cite{gard}. 
There may be situations
where neither the quasi-static nor the Langevin approaches is
appropriate. In these cases, one is obliged to face the
time-dependence of the potential head-on -- this leads to
real analytic difficulties, especially in the quantum mechanical
case, where one must abandon the notion of eigenstates, and
tackle the time-dependent Schr\"odinger equation directly. 

The analytic challenge of explicitly time-dependent mechanics is 
great, and exactly solvable cases are valuable for providing
basic insights. The harmonic potential is widely studied in many 
areas of physics, due partly to its inherent solvability.
It is also one of the most ubiquitous potentials in Nature,
due to the existence of near equilibrium states. 
It is for these reasons that the classical and quantum mechanics
of the explicitly time-dependent harmonic potential have been
studied for many years\cite{ditt,class,sch,har,hag,sik}.

One of the most extreme limits within this class of problems is that of
the pulsed harmonic potential (PHP). That is to say, the potential
exists for extremely short instants, between which there is no
potential whatsoever. This may be represented by
\begin{equation}
\label{pot}
V({\bf x},t) = v({\bf x})\sum _{n} \delta (t-\tau _{n}) \ ,
\end{equation}
where in particular 
\begin{equation}
\label{harmon}
v({\bf x}) = (\lambda /2)x^{2} \ , 
\end{equation}
and for regular pulsing $\tau _{n} = n\tau $.
(An alternative terminology has arisen in the field of quantum chaos
\cite{gutz},
in which one speaks of the system being `kicked' by such a potential.)
Apart from more obvious manifestations of such a potential (in which
an experimentalist externally pulses a system with some form of trapping
potential), one can
envisage such a situation arising in the frame of reference of a
rapidly moving particle as it regularly passes through regions within 
which a static harmonic potential exists.

To our knowledge there is no complete treatment of this system. 
In fact, we suspect that its apparent simplicity may have
persuaded workers to add complicating features. We are aware of similar
harmonic models in the field of quantum chaos\cite{berry,ford}, 
but these are typically
considered on a toroidal phase space in order to make closer contact
to classically chaotic systems. In such a case, it is known that only
periodic orbits exist in the quantum case, whereas the classical system
may be chaotic depending on the curvature of the potential. We are content
to study the system on the full phase space. Also, a great deal of effort
has been directed towards the problem of a `kicked harmonic 
oscillator'\cite{kho},
by which is meant a static harmonic potential periodically pulsed with
some spatially periodic potential. This is used to model the behaviour
of a trapped particle under the action of a laser. 

To give a unifying flavour to the present work,
we shall study three manifestations of the problem. In section
II we consider a classical particle in a PHP and parameterize the
region of stability, along with its associated periodic and
quasi-periodic dynamics. In sections III-V we consider a quantum
particle in a PHP. Sections III and IV are concerned with a wave packet
centered at the origin, which has no classical analogue. Two different
analytic formulations of the problem are presented, each with their
advantages in application. We parameterize the stability and
periodicity of the dynamics (including a bizarre cycle with period
$\tau $) and also examine the limit of $\tau \rightarrow 0$ which we
compare to the static harmonic potential. In section V we examine an
off-centered wave packet. Contact is made with the classical system
via Ehrenfest's theorem, and we also show that the expectation value
of the Hamiltonian splits neatly into two pieces which are, respectively,
purely quantum and classical in origin. In section VI we study the third 
manifestation of a PHP; namely a directed line in thermal equilibrium
with a set of planar harmonic potentials. We determine the asymptotic
transverse fluctuations of the line for binding potentials, and the
finite maximal length of the line for unbinding potentials. We end
the paper with section VII in which we give a detailed summary of our
results, along with some general conclusions.

\section{Classical Particle in PHP}

We consider a classical particle of mass $m$ in a PHP, which is
equivalent to the periodic impulsive force
\begin{equation}
\label{imp}
F(x,t) = -\lambda x \sum \limits _{n=1}^{\infty} \delta (t-n\tau) \ .
\end{equation}
The particle will suffer a discontinuous change in momentum
with a period of $\tau$. In the intervening intervals, the
particle changes its position with a constant velocity. It is
sufficient to describe the particle's trajectory by 
\begin{equation}
\label{xandp}
x_{n} \equiv \lim _{\epsilon \rightarrow 0} \ x(n\tau - \epsilon) \ , 
\hspace{1cm}
p_{n} \equiv \lim _{\epsilon \rightarrow 0} \ p(n\tau - \epsilon) \ .
\end{equation}
These quantities satisfy the difference equations:
\begin{eqnarray}
\label{itxp}
\nonumber
x_{n+1} & = & x_{n} + (\tau/ m)p_{n+1} \\
p_{n+1} & = & p_{n} - \lambda x_{n} \ .
\end{eqnarray}
We must also specify the initial conditions for the particle:
$x(0)$ and $p(0)$. The initial values for the difference
equations may then be given as: $p_{1}=p(0)$, and
$x_{1}=x(0)+(\tau/m)p(0)$.

It is convenient to rescale the momentum to $\rho _{n}=(\tau/m)p_{n}$,
so that the difference equations now take the form
\begin{eqnarray}
\label{itxr}
\nonumber
x_{n+1} & = & x_{n} + \rho_{n+1} \\
\rho_{n+1} & = & \rho_{n} - \xi x_{n} \ ,
\end{eqnarray}
where 
\begin{equation}
\label{dimcoup}
\xi = \lambda \tau /m \ ,
\end{equation}
is the dimensionless coupling to the
potential. It is a simple matter to eliminate one set of the difference
functions ($\lbrace x_{n} \rbrace$ say), to give the single
second-order difference equation
\begin{equation}
\label{diffrho}
\rho _{n+2} = \beta \rho _{n+1} - \rho _{n} \ ,
\end{equation}
where $\beta = 2-\xi$. We supply the two required
initial data $\rho _{1}$ and $\rho _{0} \equiv 
\rho _{1}+\xi (x_{1}-\rho_{1})$. [One may also proceed by recasting
Eq.(\ref{itxr}) in matrix form and determining the dynamics and stability
of the system from the associated eigenvalues. We shall use the
second-order difference equation in order to make closer contact with
the analysis of the quantum system in the following section.]

This difference equation may be easily solved by introducing the 
generating function
\begin{equation}
\label{genfn}
R(z) = \sum \limits _{n=0}^{\infty} z^{n}\rho _{n} \ ,
\end{equation}
which may be inverted via the contour integral,
\begin{equation}
\label{invgenfn}
\rho _{n} = {1\over 2\pi i} \ \int _{C} {dz \over z^{n+1}} R(z) \ ,
\end{equation}
where $C$ encircles the origin counterclockwise 
(with a radius chosen small enough so as
not to enclose any singularities bar the pole at the origin).

Summing the difference equation for $\rho $ with a weight of $z^{n}$ 
yields the following expression for the generating function:
\begin{equation}
\label{genfnex}
R(z) = {(1-\beta z)\rho _{0} + z\rho _{1} \over z^{2}-\beta z + 1} \ .
\end{equation}
This function has two simple poles located at $\alpha _{1}$ and
$\alpha _{2}$ where
\begin{equation}
\label{poles}
\alpha _{1,2} = {\beta \over 2} \pm {1\over 2}(\beta ^{2}-4)^{1/2} \ .   
\end{equation}
For $|\beta |>2$ the poles lie on the real axis and one of them has a 
modulus less than unity. Referring to Eq.(\ref{invgenfn}), we see this
implies that $|\rho _{n}|$ grows unboundedly with increasing $n$. Thus,
stable evolution of the particle is only possible for $|\beta |\le 2$
(corresponding to $0 \le \xi \le 4$), which we now examine in more detail.
Defining a parameter $\phi $ via $\beta = 2\cos \phi$, we have
$\alpha _{1,2} = e^{\pm i\phi }$.
We evaluate the contour integral in Eq.(\ref{invgenfn}) by noting
that the contour $C$ may be deformed around the singularities away
from the origin such that $\int _{C} = -\int _{\rm sing}$. In other
words, the required integral along $C$ is equal to minus the residues from
the two poles. Evaluating these residues, and performing some 
algebraic manipulations, we arrive at the result
\begin{equation}
\label{rhonex}
\rho _{n+1} = {1 \over \sin \phi } \left [ \rho _{1} \sin (n+1)\phi 
- \rho _{0} \sin n\phi \right ] \ .
\end{equation} 
It is clear from the above expression that some form of
cyclic behaviour with period $n\tau $ 
will occur when the value of $\xi $ (and thus
$\beta $) is adjusted so that $\sin n\phi = 0$. In this case we
have $\phi = M\pi /n$, where $M=1,\cdots,[n/2]$, and consequently,
$\rho _{n+k} = (-1)^{M}\rho _{k}$. From Eq.(\ref{itxr}) it is
easy to check that such a value of $\phi $ also implies
$x_{n+k} = (-1)^{M}x_{k}$. 

Let us classify two types of periodic motion: PMI($n$) -- a motion for
which all measurable quantities take on the same values with a
period of $n \tau $; and PMII($n$) -- a motion for which only the
energy of the system has a period of $n \tau $. Now, the energy
(between pulses) is simply given by 
\begin{equation}
\label{meanencl}
E_{n}={m \over 2} \left ( {\rho_{n}\over \tau}\right )^{2}
\end{equation}
Thus, PMII($n$) occurs for {\it any}
value of $\phi $ satisfying $\sin n\phi = 0$. However, PMI($n$) only
occurs for values of $\phi $ which are an {\it even} multiple
of $\pi /n$. So the simplest PMII motion occurs for $n=2$, which
corresponds to $\xi = 2$. However, the simplest PMI motion occurs
for $n=3$ with $M=2$, and corresponds to $\xi = 3$. We stress that
these periodic motions exist once the parameter $\xi $ is tuned to
an appropriate value, regardless of the initial data $(x(0),p(0))$. 

As a final remark, we note that there exists one special PMI motion
which has a period of $2\tau $. This may be seen directly from
the first order difference equations (\ref{itxr}). Such a motion is
possible if one tunes $\xi = 4$ {\it and} adjusts the initial
data such that $\rho _{1} = 2x_{1}$. 

\section{Quantum Particle in PHP: 1}

In this and the next two sections we shall examine the evolution of
a gaussian wave-packet under the influence of a PHP. We shall
highlight the similarities of the mean motion to the classical
dynamics described in section II, as well as some subtle effects
which are of a purely quantum origin. In this section we shall
examine the evolution of a wave-packet centered at the origin, and 
we shall use direct evaluation of gaussian integrals to arrive at
our results. In  the next section, we study the same problem, but
with the aid of Fourier decomposition. Both methods yield the same
results, but in surprisingly different formats, which are 
individually suited
to the calculation of different quantities. The final section of the
three is concerned with the evolution of an off-centered
wave-packet (which has non-zero expectation values for position
and momentum and may therefore be compared to the classical case).

Our starting point is Schr\"odinger's equation for the wave function
$\psi (x,t)$, with the potential given as in Eqs.(\ref{pot}) and
(\ref{harmon}):
\begin{equation}
\label{schreqn}
i\hbar \partial _{t} \psi = -{\hbar ^{2} \over 2m}\partial _{x}^{2}\psi
+ {\lambda \over 2}x^{2} \sum \limits _{n=1}^{\infty} \delta (t-n\tau )
\ \psi \ .
\end{equation}
As an initial condition we take a centered gaussian
\begin{equation}
\label{initwfn}
\psi (x,0) = (2B^{R}/\pi )^{1/4} \exp [-Bx^{2}] \ ,
\end{equation}
where $B$ is complex, and $B^{R} \equiv {\rm Re}[B] > 0$. This choice
is made on the grounds of simplicity, but we can imagine preparing
such a state from the lowest eigenstate of a static harmonic
potential. The evolution of the wave function between pulses is simply free 
particle propagation. Therefore we can describe the dynamics by the set of 
functions
$\lbrace \psi _{n}(x)\rbrace $, where 
\begin{equation}
\label{wavefnn}
\psi _{n}(x) = \lim _{\epsilon \rightarrow 0} \psi (x,n\tau - \epsilon ) \ .
\end{equation}
It is not a trivial matter to determine the change in the wave function
due to a pulsed potential. We refer the reader to Ref.\cite{new} for
a full discussion of this point 
(in the statistical mechanics context). The result is that
the wave function suffers a discontinuity in phase. Thus, the
probability density of the particle is unchanged in the immediate
temporal vicinity of the pulse.
With regard to the general pulsed
potential given in Eq.(\ref{pot}), the wave function immediately
after the pulse is given by
\begin{equation}
\label{discon1}
\psi _{n}^{+}(x) \equiv \lim _{\epsilon \rightarrow 0} 
\psi (x,n\tau + \epsilon) = \psi _{n} \exp \left [ -{iv(x) \over \hbar }
\right ] \ ,
\end{equation}
which has a very natural form when viewed from the path intregral perspective
\cite{pi}. In the present case of a PHP we have
\begin{equation}
\label{discon2}
\psi _{n}^{+}(x) = \psi _{n} \exp \left [ -{i\lambda x^{2} \over 2\hbar} 
\right ] \ .
\end{equation}
The free particle propagation between pulses may be written as
\begin{equation}
\label{fwp}
\psi _{n+1}(x) = \int dx' \ G(x-x',\tau)\psi _{n}^{+}(x') \ ,
\end{equation}
where the Green function has the familiar form\cite{schiff}
\begin{equation}
\label{greenfn}
G(x,t) = \left ( {m\over 2\pi i \hbar t} \right )^{1/2}\exp 
\left [ {imx^{2}\over 2\hbar t} \right ] \ .
\end{equation}
Combining Eqs.(\ref{discon2}) and (\ref{fwp}) yields the
iteration rule (which resembles a transfer matrix in the statistical 
mechanics context) for the functions $\lbrace \psi _{n}(x)\rbrace $; 
namely,
\begin{equation}
\label{iterpsi1}
\psi _{n+1}(x) = \int dx' \ G(x-x',\tau) \exp \left [-{i\lambda x'^{2}
\over 2\hbar} \right ] \psi _{n}(x') \ .
\end{equation} 
It is convenient to rescale $x \rightarrow y = x/b$, where
$b=(\hbar \tau/m)^{1/2}$
and define the dimensionless parameters, $\xi = \lambda \tau/m$
(cf. Eq.(\ref{dimcoup}) in section II), and $\eta = 2Bb^{2}$.
The iteration rule now takes the form
\begin{equation}
\label{iterpsi2}
\psi _{n+1}(y) = (2\pi i)^{-1/2}
\int dy' \ \exp \left [ {i\over 2}(y-y')^{2}
-{i\xi \over 2}y'^{2} \right ] \psi _{n}(y') \ ,
\end{equation} 
and from Eq.(\ref{initwfn}), the initial wave function is
given by
\begin{equation}
\label{initwfnrsc}
\psi (y,0) = (\eta ^{R}/\pi b^{2})^{1/4} \exp [-\eta y^{2}/2] \ ,
\end{equation}
where $\eta ^{R} = {\rm Re}[\eta ]$.

We shall concentrate on calculating two important physical
quantities: the probability density $P(x,t) = |\psi (x,t)|^{2}$
and the expectation value of the Hamiltonian (between pulses).
The latter is defined at the moment prior to the pulse:
\begin{equation}
\label{meanen}
E_{n} = \int dx \ \psi _{n}(x)^{*} \left [ -{\hbar ^{2}
\over 2m} \partial _{x}^{2} \right ] \psi _{n}(x) \ ,
\end{equation}
but is constant for the duration of the interval between two
adjacent pulses. 

The iteration rule (\ref{iterpsi2}) clearly shows that the wave function
will have a gaussian form for all times, given that its initial form
is chosen to be a gaussian. Thus, we write the general form
for $\psi_{n}$ (in the unscaled $x$ coordinate) as
\begin{equation}
\label{gauspsi}
\psi _{n} (x) = A_{n} \exp \left [ -\sigma _{n} {x^{2}\over 2b^{2}}
\right ] \ ,
\end{equation}
where $A_{n}$ and $\sigma _{n}$ are complex numbers.
The probability density is given by
\begin{equation}
\label{probden}
P_{n} (x) = |A_{n}|^{2} \exp [-C_{n}x^{2}/b^{2}] \ ,
\end{equation}
where $C_{n} = (\sigma _{n}+\sigma _{n}^{*})/2 = \pi b^{2}|A_{n}|^{4}$,
the latter equality following from normalization.
From Eqs.(\ref{meanen}) and (\ref{gauspsi}) one also has
\begin{equation}
\label{meanen2}
E_{n} = {\hbar \over 2\tau } \ {|\sigma _{n}|^{2} \over 
(\sigma _{n}+\sigma _{n}^{*})} \ .
\end{equation}
It is useful to define 
\begin{equation}
\label{spatint}
I_{n} \equiv \int dx \ \psi _{n}(x) = A_{n} \ 
(2\pi b^{2}/\sigma _{n})^{1/2} \ ,
\end{equation}
which may be inverted to yield
\begin{equation}
\label{finalbn}
\sigma _{n} = 2\pi b^{2} (A_{n}/I_{n})^{2} \ .
\end{equation}
Therefore, we may determine all the quantities of interest by
evaluating $A_{n}$ (the value of the wave function at the origin),
and $I_{n}$ (the spatial integral of the wave function). 

As a first step in the evaluation of these two quantities,
let us explicitly iterate the function $\psi _{n}$ back
to the initial condition. Using Eqs. (\ref{iterpsi2}) and
(\ref{initwfnrsc}) we have
\begin{equation}
\label{wfnexpl}
\psi _{n}(y) =  (\eta ^{R}/\pi b^{2})^{1/4} \ (2\pi i)^{-n/2} 
\int dy_{n-1} \cdots \int dy_{0} \ \exp \left [
-{1\over 2} y_{l}M^{(n)}_{lm}y_{m} \right ] \ ,
\end{equation}
where the $n \times n$ matrix ${\bf M}^{(n)}$ has diagonal 
elements $M^{(n)}_{00}=\eta-i$, $M^{(n)}_{ll}=-i\beta$, and 
off-diagonal elements $M^{(n)}_{lm}=i$ for $|l-m|=1$, and
$M^{(n)}_{lm}=0$ otherwise.

Thus, the wave function at the origin is given by
\begin{equation}
\label{an1}
A_{n} =  (\eta ^{R}/\pi b^{2})^{1/4} \ i^{-n/2} \ q_{n}^{-1/2} \ ,
\end{equation}
where $q_{n} \equiv {\rm det} \ {\bf M^{(n)}}$.
Also, integrating the above $n$-fold integral over $y$,
we find 
\begin{equation}
\label{in1}
I_{n} = (2\pi b^{2})^{1/2}
(\eta ^{R}/\pi b^{2})^{1/4} \ i^{-n/2} \ (-iq_{n} + q_{n-1})^{-1/2} \ .
\end{equation}
So we may describe the entire dynamics from the set of determinants
$\lbrace q_{n} \rbrace $. Before explicitly calculating these
functions, we shall first express the physical quantities of interest
in terms of $\lbrace q_{n} \rbrace $.
From Eq.(\ref{finalbn}) we may combine the above two expressions
to give (for $n \ge 1$)
\begin{equation}
\label{bn1}
\sigma _{n} = -i \left ( {q_{n} + iq_{n-1} 
\over q_{n} } \right ) \ .
\end{equation}
We therefore have the explicit form for the wave function (for $n \ge 1$):
\begin{equation}
\label{wavefn1}
\psi _{n}(x) = (\eta ^{R}/\pi b^{2})^{1/4} \ i^{-n/2} \ q_{n}^{-1/2}
\exp \left \lbrace {i x^{2} \over 2b^{2}} \left ( {q_{n} + iq_{n-1} 
\over q_{n} } \right ) \right \rbrace \ .
\end{equation}

The wave function at intervening times may be easily found by
propagating the above form with the Green function (\ref{greenfn}).
Straightforward integration yields 
\begin{eqnarray}
\label{wavefn2}
\nonumber
\psi (x,(n-1)\tau + \theta \tau) = (\eta ^{R}/\pi b^{2})^{1/4} \ 
i^{-n/2} & & [\theta q_{n}-i(1-\theta q_{n-1}]^{-1/2}\\
& & \times \exp \left \lbrace {ix^{2} \over 2b^{2}} 
\left [ {q_{n} + iq_{n-1} 
\over \theta q_{n} - i(1-\theta q_{n-1}) } \right ] \right \rbrace \ ,
\end{eqnarray}
where $\theta \in [0,1]$.

We may determine the probability density just prior to pulsing (
{\it i.e.} the function $P_{n}(x)$) either from our knowledge of $A_{n}$
(cf. Eq.(\ref{probden})), or by simply taking the modulus squared
of $\psi _{n}$ as given in Eq.(\ref{wavefn1}). The results of these
procedures, although strictly identical, are not obviously so, since
their equivalence requires the following identity to hold:
\begin{equation}
\label{qiden}
{\rm Re}[q_{n}q_{n-1}^{*}] = \eta ^{R} \ .
\end{equation}
The proof of this statement will be given shortly. The explicit form
of the probability density is given by (for general times)
\begin{equation}
\label{probden1}
P(x,(n-1)\tau + \theta \tau) = 
{(\eta ^{R}/\pi b^{2})^{1/2} \over |\theta q_{n} -i (1-\theta )q_{n-1}|} \
\exp \left \lbrace -{x^{2}\over b^{2}} \ {\eta ^{R} \over   
|\theta q_{n} -i (1-\theta )q_{n-1}|^{2}}  \right \rbrace \ .
\end{equation}
The important information contained in the probability density is
the temporal evolution of the width of the gaussian wave-packet.
Just prior to pulsing we have $P_{n} \sim \exp [-x^{2}/\gamma _{n}^{2}]$,
where the width $\gamma _{n}$ is simply given by 
\begin{equation}
\label{width}
\gamma _{n} = |q_{n}|\gamma _{0} \ ,
\end{equation}
where the initial width $\gamma _{0}=b/\sqrt{\eta ^{R}}$.
Finally, for the expectation value of the Hamiltonian between pulses
({\it i.e.} the mean energy), we may derive from Eqs.(\ref{meanen2}),
(\ref{bn1}) and (\ref{qiden}) the result
\begin{equation}
\label{meanen3} 
E_{n} = {\hbar \over 4\tau \eta ^{R}} \ |q_{n} + iq_{n-1}|^{2} \ .
\end{equation}

We now turn to the evaluation of the set of determinants  
$\lbrace q_{n} \rbrace $. From the definition of the matrix
${\bf M}^{(n)}$ it is straightforward to derive the following
iteration rule
\begin{equation}
\label{diffqn}
q_{n+2} = -i\beta q_{n+1} + q_{n} \ ,
\end{equation}
where $\beta = 2-\xi$. This is very similar to the second-order difference
equation (\ref{diffrho}) that 
we derived previously for the rescaled momenta in the
classical system. In the present case, the initial data 
are $q_{0}=1$, and $q_{1}=\eta - i$. Before proceeding to solve
Eq.(\ref{diffqn}), let us first prove the assertion (\ref{qiden}).
Multiplying through the difference equation (\ref{diffqn}) by
$q_{n+1}^{*}$ yields
\begin{equation}
\label{diffqn1}
q_{n+2}q_{n+1}^{*} = -i\beta |q_{n+1}|^{2} + q_{n+1}^{*}q_{n} \ .
\end{equation}
Now the first term on the right-hand-side is purely imaginary, thus
the bilinear object ${\rm Re}[q_{n}q_{n-1}^{*}]$ is independent of $n$
and therefore equal to 
\begin{equation}
\label{qiden1}
{\rm Re}[q_{n}q_{n-1}^{*}] = {\rm Re}[q_{1}q_{0}^{*}] = {\rm Re}[\eta -i] = 
\eta ^{R} \ ,
\end{equation}
as required.

The solution of the difference equation (\ref{diffqn})
may be found by use of a generating function just as in
the classical system. Bounded evolution of $|q_{n}|$ is again limited
to the parameter regime $|\beta | \le 2$, which corresponds to 
$0 \le \xi \le 4$. It is convenient to define the parameter $\phi $ via
\begin{equation}
\label{phieta}
\cos \phi = \beta /2 = 1 - \xi /2 \ , 
\end{equation}
as in the classical system. 
The explicit form for $q_{n}$ is found to be:
\begin{equation}
\label{qnexp}
q_{n} = {(-i)^{n} \over \sin \phi } \left [ i(\eta -i) \sin n\phi
-\sin (n-1)\phi \right ] \ .
\end{equation}
It is very tempting to make a direct connection between these results,
and those for the classical system. However, in doing so we would
lose some of the subtleties contained within. Consider first, the
present quantum dynamics has no classical analogue, since the expectation
values of the position and momentum remain zero for all times.
Second, there is a difference between the types of periodic motion
in the quantum system considered here, and the classical system.
In the latter we found it useful to classify periodic behaviour into
two categories: PMI($n$) (all physical quantities having period $n\tau $) and
PMII($n$) (the energy having period $n\tau $). The condition for motion
of the first class was $\phi = M\pi /n$, with $M$ even; whereas the
condition for the second class was $\phi = M\pi /n$ with $M$ odd or even.
In the present case of quantum motion, it is clear from Eq.(\ref{qnexp})
that periodic behaviour
is to be expected for $\sin n\phi = 0$ which corresponds to 
$\phi = M\pi /n$ with $M=1,\cdots,[n/2]$. However, apart from setting the
scale of $\phi$, the integer $M$ plays no other role, since it appears
in the wave function as a constant phase factor $e^{-iM\pi/2}$. So there
is no distinction between PMI and PMII motions in the quantum dynamics
of a centered gaussian wave function. 

It is of interest to compare the
energy portraits for the classical and quantum motions. We scan through
values of $\xi \in (0,2)$ and follow the evolution of the energy 
(as given by Eqs.(\ref{meanencl}) and (\ref{meanen3}) respectively) for
twenty or so iterations, all of which values are plotted. We refer the reader
to Figs.1 - 5 for the classical (with typical values of $x(0)$ and
$p(0)$) and quantum (with varying initial parameter $\eta $)
portraits. Despite the similarity in the values of 
$\xi$ at which the periodic orbits occur, we see a distinct
difference in the bounding curves $E_{\rm min}(\xi )$ and
$E_{\rm max}(\xi )$ for the minumum and maximum energies.
In the classical case, $E_{\rm min}$ is essentially zero and $E_{\rm max}$
is a monotonically increasing function of the coupling $\xi $. In the
quantum case, $E_{\rm min}$ maintains a distinct gap from zero, and
the difference $E_{\rm max}-E_{\rm min}$ is non-monotonic with $\xi $. It
would be of interest to determine the analytic properties of these bounding 
curves, but this is beyond the scope of the present work.

We shall make a more physical connection between the quantum and classical
motions in section V when we consider an off-centered wave function which
allows the particle to maintain non-zero expectation values of both
position and momentum. First however, we present an alternative treatment
of the centered gaussian wave function, which will allow us to study some
other properties of this system more easily.

\section{Quantum Particle in PHP: 2}

Although the calculations of the previous section were reasonably
straightforward, they still required the evaluation of cumbersome
$n$-fold gaussian integrals. It might be hoped that a simpler 
derivation of the results is possible, since the wave function
evolves by free wave propagation between pulses, which may
be handled more easily in Fourier space. The purpose of this section
is to derive an iteration rule for the wave function using this
Fourier method. The surprise is that the rule (or difference equation)
turns out to be non-linear and of first order, in contrast to the
linear second order rule (\ref{diffqn}) derived above. The two 
iteration rules must yield the same results of course; but, as we shall
see, they are non-trivially related. 

To proceed, we restate that given our initial wave function is gaussian,
we can expect the wave function at all subsequent times to retain
a gaussian form. So, we write
\begin{equation}
\label{gauspsi1}
\psi _{n}(x) \sim \exp \left [ -\sigma _{n} {x^{2} \over 2b^{2}}
\right ] \ ,
\end{equation}
where we have omitted the prefactor. 
The wave function at the instant after the pulse is given by 
Eq.(\ref{discon2}) which we rewrite here as
\begin{equation}
\label{gauspsi2}
\psi _{n}^{+}(x) \sim \exp \left [ - (\sigma _{n} + i\xi)          
{x^{2} \over 2b^{2}} \right ] \ .
\end{equation}
The next stage of the evolution is free wave propagation, which is
most easily expressed in Fourier space: ${\tilde \psi }(k,t+\tau) =
{\tilde \psi }(k,t)\exp [-ib^{2}k^{2}/2]$. Now the Fourier transform
of $\psi _{n}^{+}$ takes the form
\begin{equation}
\label{gauspsi3}
{\tilde \psi} _{n}^{+}(k) \sim
\exp \left [-{k^{2}b^{2}\over 2(\sigma _{n} + i\xi)} \right ] \ ,
\end{equation}
and so
\begin{equation}
\label{gauspsi4}
{\tilde \psi} _{n+1}(k) \sim
\exp \left [-{k^{2}b^{2}\over 2} \left ({1\over \sigma _{n} + i\xi }
+ i \right ) \right ] \ .
\end{equation}
Finally, inverse transforming the above expression we arrive at
\begin{equation}
\label{gauspsi5}
\psi _{n+1}(x) \sim \exp \left [ -\sigma _{n+1} {x^{2} \over 2b^{2}}
\right ] \ ,
\end{equation}
where
\begin{equation}
\label{nonlinit}
{1 \over \sigma _{n+1}} = {1 \over \sigma _{n} + i\xi } + i \ .
\end{equation}
This iteration rule requires only one piece of initial datum. Evolving the
initial wave function to $t=\tau - 0$, we have
\begin{equation}
\label{initsig}
\sigma _{1} = \eta /(1+i\eta ) \ .
\end{equation}

Given this non-linear iteration rule, one might expect the system to
have some kind of non-trivial (chaotic) dynamics. However, with the 
benefit of hindsight, we know from section III that in fact the system
has either periodic or quasi-periodic behaviour (in the stable regime
$0 \le \xi \le 4$). This simpler behaviour is not apparent from the 
non-linear rule for $\lbrace \sigma _{n} \rbrace$, 
but we can prove it is the case by connecting this
rule to the linear second order rule for $\lbrace q_{n} \rbrace$ given
in Eq.(\ref{diffqn}).

Referring to (\ref{bn1}), we have the relation (for $n \ge 1$)
\begin{equation}
\label{relqsig}
\sigma _{n} = -i \left ( {q_{n} + iq_{n-1} \over q_{n}} \right )  \ .
\end{equation}
Substituting this result into the non-linear iteration rule
(\ref{nonlinit}), we find equality only if $q_{n}$ satisfies the
linear iteration rule (\ref{diffqn}), as required for consistency. 
In the absence of hindsight, the relation (\ref{relqsig}) would
be seen as a rather remarkable linearizing transformation. 
In order to probe the general (quasi)-periodic behaviour of this system,
the linear iteration rule is the description of choice. However, there
are two other aspects of this system which are much more easily
described by the non-linear rule given above.

The first of these is the existence of a special periodic motion with
period $\tau $. We may identify this by simply demanding that
$\sigma _{n+1} = \sigma _{n} \ (= \sigma )$ in Eq.(\ref{nonlinit}).
This yields a quadratic equation for $\sigma $ with (normalizable) solution
\begin{equation}
\label{quadsol}
\sigma = {1 \over 2} [ \xi (4-\xi )]^{1/2} - {i\xi \over 2} \ .
\end{equation}
Now, the initial value of $\sigma $ is set by Eq.(\ref{initsig}). So
this special `1-cycle' is only possible (for a given value of
$\xi \in [0,4]$) for a special value ${\bar \eta}$ of the parameter 
$\eta $ (which, we remind the reader, describes the initial gaussian
wave function).  
We may invert (\ref{initsig}) for ${\bar \eta} $ to find
\begin{equation}
\label{etasol}
{\bar \eta } \ = \ \sin \phi + i(1-\cos \phi ) \ = \
2\sin (\phi /2 ) \exp (i\phi /2) \ ,
\end{equation}
where we have used $\phi $ in place of $\xi $, as defined in 
Eq.(\ref{phieta}). An example of this 1-cycle is shown in Fig. 5.
This 1-cycle is a purely quantum effect, since the classical system
can only have cycles of period $\ge 2\tau $ for obvious reasons.
One might term this motion `the sound of one hand clapping.'

The second aspect of the quantum system which may be extracted
more easily from the non-linear iteration is the behaviour of the
system as the time between pulses is taken to zero. Before taking this
limit, it is important to scale out $\tau $ in other quantities.
Now, we shall compare the system (in the limit of $\tau \rightarrow 0$),
to the quantum mechanics of the static
harmonic potential. For the latter system, the potential is taken to 
be $V_s(x) = \kappa x^{2}/2 $. The time-averaged potential
of the PHP is ${\bar V}(x) = \lambda x^{2}/2\tau $. Thus we shall take the 
limit $\tau \rightarrow 0$ and $\lambda \rightarrow 0$ with the ratio
$\lambda /\tau \rightarrow \kappa $. It is also convenient to
define $v_{n}=\sigma _{n}/\tau $. Then the wave function just prior
to pulsing is given by (cf. Eq.(\ref{gauspsi1}))
\begin{equation}
\label{wavefnm}
\psi _{n}(x) \sim \exp \left [ -{m \over 2\hbar} v_{n}x^{2} \right ] \ .
\end{equation}
From (\ref{nonlinit}), the iteration rule for $v_{n}$ takes the form
\begin{equation}
\label{nonlinitv}
{1 \over v _{n+1}} = {1 \over v_{n} + i\lambda /m } + i\tau \ .
\end{equation}
Taking the $\tau \rightarrow 0$ limit as prescribed above, the above
iteration rule tends to the first-order differential equation for $v(t)$:
\begin{equation}
\label{diffv}
{dv \over dt} = i\left ( {\kappa \over m} - v^{2} \right ) \ .
\end{equation}

Now, the ground state wave function of the static harmonic potential
is a simple gaussian $\varphi (x) \sim \exp (-m\omega _{s}x^{2}/2\hbar )$,
where the oscillator frequency $\omega _{s}= \sqrt{\kappa /m}$. Referring
to the differential equation for $v(t)$, we see that $v=\omega _{s}$ is
a fixed point of the dynamics. In other words, if we initialize the
wave function to be the ground state wave function of the static
harmonic potential, then the wave function will be completely
unaffected by the PHP in the limit of $\tau \rightarrow 0$. We stress
that there will be no dynamical evolution whatsoever. This is in contrast
to the wave function of a truly static harmonic potential, which 
if prepared in the ground state, will still have a dynamically
evolving phase factor $e^{-i\omega _{s}t/2}$. The pulsing of the potential 
essentially resets the clock of the wave function such that the
dynamical phase is `stuck' at $t=0$. This effect may have important
consequences for numerical integration of the time-dependent
Schr\"odinger equation. If one places the equation on a discrete
temporal grid, then one is pulsing the potential. From the present
example, we see that a pulsed potential (on however fine a grid)
cannot mimic a static potential. The dynamical phase information
is irretrievably lost. Subjecting a system to very high frequency 
pulses was also studied recently in the context of controlling 
decoherence\cite{vit}.

One can examine the dynamics of the PHP via the differential equation
(\ref{diffv}) in more detail. For instance, one can examine the
evolution of a gaussian wave packet which is not tuned to be 
$\varphi (x)$. Let us restrict ourselves to an initial condition
for $\psi $ which is real, and therefore completely described by the width of
the wave packet, which we denote by $l_{0}$, and which is related to
an effective frequency $\omega = \hbar/ ml_{0}^{2}$. 
A straightforward solution of (\ref{diffv}) yields
\begin{equation}
\label{psisol}
\psi (x,t) \sim \exp \left [ -{x^{2} \over 2l_{0}^{2}} \left (
{1+(i/r) \tan (\omega t) \over 1+ir\tan (\omega t) }\right ) \right ] \ ,
\end{equation}
where $r=\omega/\omega _{s}$.
Consequently, the probability density $P(x,t) \sim \exp (-x^{2}/\gamma ^{2})$
with 
\begin{equation}
\label{probwid}
\gamma (t) = l_{0} \left ( {1+ r^{2} \tan ^{2}(\omega t) \over 
1+\tan ^{2} (\omega t) }\right ) \ .
\end{equation}
The expectation value of the kinetic energy is given by
\begin{equation}
\label{kinen}
E_{\rm kin} (t) = {\hbar \omega \over 4}
\left ( {1+ (1/r)^{2} \tan ^{2}(\omega t) \over 
1+\tan ^{2} (\omega t) }\right ) \ .
\end{equation}
In the limit of $\tau \rightarrow 0$ there is also an effective
potential energy (from averaging over the many pulses in a given
small time interval). The expectation value of the potential energy
may be found from $\int dx \psi ^{*} (\kappa x^{2}/2)\psi$, with
$\psi (x,t)$ given by (\ref{psisol}) above. One finds
\begin{equation}
\label{poten}
E_{\rm pot} (t) = {\hbar \omega \over 4r^{2}}  
\left ( {1+ r^{2} \tan ^{2}(\omega t) \over 
1+\tan ^{2} (\omega t) }\right ) \ .
\end{equation}
The total energy is then found to be 
\begin{equation}
\label{toten}
E_{\rm tot} (t) = {\hbar \omega _{s} \over 4} \left (r^{2}+{1\over r^{2}}
\right ) \ ,
\end{equation}
which is independent of time as expected.

\section{Quantum Particle in PHP: 3}

In this third and final section concerning quantum dynamics
in a PHP, we shall consider an initial wave function which is 
off-centered. Thus the expectation values of the position and
momentum of the particle will be non-zero, and we may make
closer contact between the quantum PHP and its classical
counterpart. We shall utilize the iteration
rules from both sections III and IV.

We begin by deriving the first order iteration rules using
the method described in section IV. Consider an off-centered
gaussian at the moment prior to the $n^{\rm th}$ pulse:
\begin{equation}
\label{offcen}
\psi _{n} \sim \exp \left [ -\sigma _{n} {x^{2} \over 2b^{2}}
+ d_{n}{x \over b} \right ] \ .
\end{equation}
Following the treatment for the centered gaussian (cf
Eqs.(\ref{gauspsi1}) - (\ref{gauspsi5})) we find
that the iteration rule for $\lbrace \sigma _{n} \rbrace $
is unchanged from the form given in Eq.(\ref{nonlinit}),
and the rule for $\lbrace d_{n} \rbrace$ is
\begin{equation}
\label{itford1}
d_{n+1} = d_{n} \ {\sigma _{n+1} \over (\sigma _{n} + i\xi )} \ .
\end{equation}
With the aid of (\ref{nonlinit}) and (\ref{relqsig}) we may rewrite 
this as
\begin{equation}
\label{itford3}
d_{n+1}q_{n+1} = -i d_{n} q_{n} \ ,
\end{equation}
which may be iterated immediately to give the solution
\begin{equation}
\label{itford4}
d_{n+1} = (-i)^{n} {d_{0} \over q_{n}} \ ,
\end{equation}
where $d_{0}$ is the parameter introduced to describe the
initial off-centered wave function:
$\psi (x,0) \sim \exp [-\eta x^{2}/2b^{2} + d_{0} x/b] $.

Our present interest will not so much be in $\lbrace \sigma _{n} \rbrace $ 
and $\lbrace d_{n} \rbrace $, but rather in the expectation
values of the position and momentum. Explicitly evaluating
the appropriate expectation values using Eq.(\ref{offcen}) we find
\begin{equation}
\label{expx}
{\bar x}_{n} \equiv \int dx \ \psi ^{*} x \psi = b {(d_{n} + d_{n}^{*}) 
\over (\sigma _{n} + \sigma _{n}^{*})} \ ,
\end{equation}
and
\begin{equation}
\label{expp}
{\bar p}_{n} \equiv \int dx \ \psi ^{*} (-i\hbar \partial _{x}) 
\psi = -{i\hbar \over b} {(\sigma ^{*}d_{n} - \sigma _{n}d_{n}^{*}) 
\over (\sigma _{n} + \sigma _{n}^{*})} \ .
\end{equation}

It is convenient to re-express the above relations in terms of
$\lbrace q_{n} \rbrace $ and $\lbrace d _{n} \rbrace $; which, 
due to Eq.(\ref{itford4}), amounts to expressing the expectation
values purely in terms of $\lbrace q_{n} \rbrace $ and $d_{0}$.
Using the known relations (\ref{qiden}),(\ref{relqsig}) and
(\ref{itford4}), we have
\begin{equation}
\label{expx1}
{\bar x}_{n} = {i^{n}b\over 2 \eta ^{R}} [q_{n}d_{0}^{*} +
(-1)^{n}q_{n}^{*}d_{0}] \ ,
\end{equation}
and
\begin{equation}
\label{expp1}
{\bar p}_{n} = {i^{n}\hbar \over 2b\eta ^{R}}
[q_{n}d_{0}^{*} + (-1)^{n}q_{n}^{*}d_{0} + iq_{n-1}d_{0}^{*}
-i (-1)^{n}q_{n-1}^{*}d_{0} ] \ .
\end{equation}

We are now in a position to utilize the linear iteration rule
(\ref{diffqn}) for the $\lbrace q_{n} \rbrace $. Taking neighbouring 
differences between the expectation values of the position and
momenta, we may use (\ref{diffqn}) and the definition 
$b=(\hbar \tau /m)^{1/2}$ to derive
\begin{equation}
\label{expxit}
{\bar x}_{n+1} - {\bar x}_{n} = {\tau \over m} {\bar p}_{n+1} \ ,
\end{equation}
and 
\begin{equation}
\label{exppit}
{\bar p}_{n+1} - {\bar p}_{n} = -\lambda {\bar x}_{n} \ .
\end{equation}
A direct comparison between these iteration rules with those 
given in Eq.(\ref{itxp}) for the classical system reveals that
the {\it mean motion} of the off-centered gaussian wave function 
in the PHP is identical to the purely classical motion -- a 
non-trivial example of Ehrenfest's Theorem\cite{schiff}.

Finally, we shall consider the expectation value of the energy.
Using the form for the wave function given in Eq.(\ref{offcen})
we have (cf Eq.(\ref{meanen}))
\begin{equation}
\label{menoc}
E_{n} = {\hbar \over 2\tau } \left [ {|\sigma _{n}|^{2} \over
(\sigma _{n} + \sigma _{n}^{*})} - {(\sigma _{n}^{*}d_{n}
-\sigma _{n}d_{n}^{*})^{2} \over (\sigma _{n} + \sigma _{n}^{*})^{2}}
\right ] \ . 
\end{equation}
Referring to Eq.(\ref{meanen2}), we see that the first term on the 
right-hand-side is
precisely the `quantum' energy for the centered gaussian,
studied in section III. Referring to Eq.(\ref{expp}), the second
term on the right-hand-side is determined as nothing more than
the classical energy ${\bar p}_{n}^{2}/2m$ studied in section II. 
Thus, the energy of the off-centered gaussian
falls neatly into two separate pieces: a `classical' energy determined
purely by the expectation value of the momentum, and a `quantum'
energy determined purely by the internal fluctuations of the wave
packet in the co-moving frame. This ends the study of a quantum particle
in a PHP.

\section{Directed Polymer in PHP}

Quantum mechanical Green functions may be rewritten as path integrals.
It is within this formalism that one may appreciate the
close mathematical connection between quantum processes and
the statistical mechanics of directed lines. We shall use
this connection to discuss the physics of a directed
line in thermal equilibrium with a PHP, which is physically
realized by a set of static, planar (or sparse\cite{new}), harmonic 
potentials.

We begin by writing the Feynman path integral\cite{pi} for the Green function
of a quantum mechanical particle in a potential $V(x,t)$ (where we restrict
our attention to one dimension for simplicity):
\begin{equation}
\label{feynpi}
G(x,t;x_{0},t_{0}) = \int \limits _{y(t_{0})=x_{0}}^{y(t)=x} {\cal D}y(s) \ 
\exp \left \lbrace {i \over \hbar} \int \limits _{t_{0}}^{t} ds \ 
\left [ {m\over 2} \left ({dy \over ds} \right )^{2} - V(y(s),s)
\right ] \right \rbrace \ .
\end{equation}
This is to be compared with the (restricted) partition function
for a directed line in thermal equilibrium with a static potential
$V(x,z)$. By `directed line' we mean a connected path
in a two dimensional space (x,z),
which may adopt any configuration whatever, so long as it is directed
along the longitudinal ($z$) direction. The statistical mechanics of
such objects is of interest in several fields, including directed 
polymers\cite{dp}, 
and superconducting flux lines\cite{blat}. The partition function for 
a line with one end pinned at $(x_{0},z_{0})$ and the other pinned
at $(x,z)$ is given by
\begin{equation}
\label{dirppi}
Z(x,z;x_{0},z_{0}) = \int \limits _{y(z_{0})=x_{0}}^{y(z)=x} {\cal D}y(s) \ 
\exp \left \lbrace -{1 \over T} \int \limits _{z_{0}}^{z} ds \ 
\left [ {\kappa \over 2} \left ({dy \over ds} \right )^{2} + V(y(s),s)
\right ] \right \rbrace \ ,
\end{equation}
where $T$ is temperature (with Boltzmann's constant set to unity) and 
$\kappa $ is the elasticity of the line.

The obvious similarity between these two path integrals can be
misleading, since it is important to remember that the relation
between them exists at a strictly mathematical level. Given
the analytic solution to one path integral, one may infer the 
solution to the other by an analytic continuation from `real
time' $t$ to `imaginary time' $z$. However, the physical 
properties of the two systems are generally quite distinct
and little may be inferred about the physics of
one of the systems, if only qualitative information about the
physics of the other is available. This will become clear
in the present case of a PHP, as we shall soon see. 

The physical meaning of a directed line in equilibrium with
a PHP is as follows. The PHP itself consists of potentials 
which only exist on discrete transverse lines, and are regularly
spaced along the longitudinal axis. These potentials are harmonic
and centered at $x=0$. The directed line (which we shall take to
be pinned at $(0,0)$ and to have a length $z$) equilibrates itself
in these potentials, meaning that the free energy is minimized 
as a result of the competition between the wandering of the line
(entropy), and its elastic and potential energies. 
One may find an application within the field of superconductivity. In
an array of flux lines, one may construct an approximate (harmonic) `caging' 
potential for a given flux line, by averaging over the repulsive 
line-line interactions which it experiences with its neighbours. Furthermore,
in the strongly layered cuprates\cite{cup} (which form the most important
class of high temperature superconductors) the supercurrents
only exist in well separated Cu-O planes. Thus the flux line
(meaning the imaginary line connecting the planar centres of
magnetic flux) will only experience the caging potential in
discrete, but regular, transverse planes.

Given the relationship between the quantum system and the directed
line at the level of path integrals, there is naturally a
partial differential equation for $Z$ corresponding to the 
Schr\"odinger equation. Defining a `rigidity' $\nu = T/2\kappa $,
and absorbing $T$ into the potential, we have

\begin{equation}
\label{pfeqn}
\partial _{z}Z = \nu \partial _{x}^{2}Z - V(x,z)Z \ . 
\end{equation}
The potential is taken to be a PHP, expressed as
\begin{equation}
\label{potdp}
V(x,z) = {gx^{2}\over 2} \sum \limits _{n=1}^{\infty }\delta (z-nd) \ .
\end{equation}
The initial condition implicit in the path integral (\ref{dirppi})
is $Z(x,0)=\delta (x)$, but we may generalize this to any desired
function. Following our earlier work on the quantum system, we shall
take a gaussian initial condition
\begin{equation}
\label{initdp}
Z(x,0) = \exp [-x^{2}/a^{2}] \ ,
\end{equation}
which would naturally arise from thermal wandering of a directed line
from a $\delta $-function initial condition.
The normalization of $Z$ deserves mention. Whereas in the quantum
system, the wave function $\psi $ is normalized by requiring that
$\int dx |\psi |^{2} = 1$, the partition function has an arbitrary prefactor,
as we only require that $\int dx {\cal P}(x,z) = 1$, where 
${\cal P}(x,z)=Z(x,z)/\int dx'Z(x',z)$ is the probability density
of the line.

We shall not enter into any details concerning the analysis
of this system, as our results may be easily reconstructed from
the methods presented in section III for the quantum analogue.
The present system is described by two dimensionless parameters:
an effective coupling 
\begin{equation}
\label{couplingdp}
{\tilde g} = 2\nu d g \ ,
\end{equation}
and the ratio 
\begin{equation}
\label{ratiodp}
\chi = 4\nu d /a^{2} = (l/a)^{2} \ , 
\end{equation}
which is the square of the ratio of the transverse thermal wandering 
scale $l=(4\nu d )^{1/2}$ and the initial transverse scale $a$.
The probability density of the line just prior to the $n^{\rm th}$
impulse is defined as 
\begin{equation}
\label{probdpdef}
{\cal P}_{n}(x) = \lim _{\epsilon \rightarrow 0}
{\cal P}(x,nd - \epsilon ) \ ,
\end{equation}
and has the explicit form
\begin{equation}
\label{probdp}
{\cal P}_{n}(x) = \left [ {(1-q_{n-1}/q_{n})\over \pi l^{2}}
\right ] ^{1/2} \exp \left [ - \left ( {q_{n} - q_{n-1} \over q_{n}} 
\right ) \left ( {x \over l } \right )^{2} \right ] \ . 
\end{equation}
This gaussian form is completely described by one quantity; namely,
the width $\gamma _{n}$ of the probability density defined via
${\cal P}_{n} \sim \exp [-x^{2}/\gamma _{n}^{2}] $.
Thus
\begin{equation}
\label{widthdp}
\gamma _{n} = l \left ( {q_{n} \over q_{n} - q_{n-1}} \right )^{1/2} \ .
\end{equation}
The determinants $\lbrace q_{n} \rbrace $ satisfy the second order
difference equation
\begin{equation}
\label{itdpqn}
q_{n+2} = \beta q_{n+1} - q_{n} \ ,
\end{equation}
where $\beta = {\tilde g} + 2$, and the initial data are 
$q_{0}=1$ and $q_{1}=\chi+1$.

Let us first study the case of $\beta > 2$, which corresponds to 
an attractive (or binding) PHP with ${\tilde g}>0$. In this case
it is convenient to define a parameter $\theta $ via
\begin{equation}
\label{thetadp}
\cosh \theta = \beta /2 = 1 + {\tilde g} \ .
\end{equation}
Then we find
\begin{equation}
\label{solqndp}
q_{n} = {1\over \sinh \theta } \ [ \sinh (n+1)\theta + 
(\chi + 1 - 2\cosh \theta )\sinh n\theta ] \ .
\end{equation}
Substituting this solution into Eq.(\ref {widthdp}) yields the
final result for the transverse line scale as a function of
$n=z/d $ :
\begin{equation}
\label{widthsoldp}
\gamma _{n} = l \left [ { (\chi + 1 - \cosh \theta )\tanh n\theta \ \ + \ \ 
\sinh \theta \over
(\cosh \theta -1)(2\cosh \theta - \chi)\tanh n\theta + 
(\chi - 2(\cosh \theta -1))\sinh \theta } \right ]^{1/2} \ .
\end{equation}

For the limit of an infinitely long line, $n \theta \rightarrow \infty$, 
the function $\tanh n\theta \rightarrow 1$, and the above
result simplifies dramatically to
\begin{equation}
\label{wsoldp2}
\gamma _{\infty} = {l \over \sqrt{\zeta } }  \ ,
\end{equation}
with finite $n$ corrections $\sim e^{-2n\theta }$. The dimensionless
parameter $\zeta $ is given by
\begin{equation}
\label{zetadp}
\zeta \equiv 1 + \sinh \theta - \cosh \theta = 1 - e^{-\theta} \ .
\end{equation}
Note that for 
\begin{equation}
\label{zetah}
\zeta = \chi
\end{equation}
the transverse scale is asymptotically
equal to the initial scale $a$, meaning that a specially tuned
PHP can exactly compensate the transverse wandering for arbitrarily
long lines. The tuned value of the coupling ${\tilde g}$ needed to
satisfy Eq.(\ref{zetah}) is
\begin{equation}
\label{tuned}
{\tilde g} = {\chi^{2} \over 1 - \chi } \ . 
\end{equation}
Since the coupling ${\tilde g}$ must be positive, we see that such
a compensating PHP is only possible for $\chi <1$, {\it i.e.}
for a line whose initial
scale $a > l$. This is clear since whatever the PHP strength, the
line is free to wander a transverse scale $l$ between pulses, and thus
we can never restrict the line to $\gamma _{\infty} = a$ if $a<l$.

Regardless of the initial transverse scale $a$, a very strong PHP 
is expected to strongly compress the line to have a scale $\gamma _{\infty}
\sim l$. 
For ${\tilde g} \gg 1$, one can perform an asymptotic expansion on 
Eq.(\ref{wsoldp2}) to find 
\begin{equation}
\label{widbigg}
\gamma _{\infty} = l
\left ( 1 + {1 \over 4{\tilde g}} + \cdots \right ) \ .
\end{equation}

The case of a repulsive (or unbinding) PHP is a little more subtle
as the line will become unstable (meaning the probability density
becomes unnormalizable) beyond a critical length depending on the
strength of the potential. In fact, it is easy to see that for the 
line to survive just one pulse, we require ${\tilde g} > -2$.
It is therefore convenient to define a parameter $\theta '$ via
\begin{equation}
\label{dpthetap}
\cos \theta ' = 1 - |{\tilde g}|/2 \ .
\end{equation}
We find for the determinants
\begin{equation}
\label{detdp}
q_{n} = {1\over \sin \theta '} \ [ \sin (n+1)\theta ' + 
(\chi + 1 - 2\cos \theta ')\sin n\theta '] \ ,
\end{equation}
and using Eq.(\ref{widthdp}), 
\begin{equation}
\label{whsoldp}
\gamma _{n} = l \left [ { (\chi + 1 - \cos \theta ')\tan n\theta ' \ \ 
+ \ \ \sin \theta ' \over
(\cos \theta ' - 1)(2\cos \theta ' - \chi )\tan n\theta ' + 
(\chi - 2(\cos \theta ' -1))\sin \theta ' } \right ]^{1/2} \ .
\end{equation}
This expression is not valid for arbitrarily large $n$. There
exists a maximum length $n^{*}d $ for the directed line,
beyond which it no longer exists as a connected elastic structure.
The value $n^{*}$ may be found by demanding that $q_{n} > 0$ for
all $n \le n^{*}$. Referring to Eq.(\ref{detdp}) we find that
$n^{*} = [w^{*}]$, where 
\begin{equation}
\label{upperlen}
w^{*} = {1\over \theta '} \ \left [ {\pi \over 2} + \tan ^{-1}
\left ( {\chi+ 1 - \cos \theta '\over \sin \theta ' } \right )
\right ] \ .
\end{equation}
For $|{\tilde g}| \rightarrow 2$, $\theta ' \rightarrow \pi/2$ and
$1 < w^{*} < 2$ as expected. 

The more interesting limit of $|{\tilde g}| \rightarrow 0$ yields the result
\begin{equation}
\label{largew}
w^{*} \sim {\pi \over |{\tilde g}|^{1/2}} - {1 \over \chi } + 
O(|{\tilde g}|^{1/2}) \ .
\end{equation}
The transverse scale of the density may also be studied in the
limit of small $|{\tilde g}|$. Referring to Eq.(\ref{whsoldp}) we
find $\gamma _{n} \sim l\sqrt {n}$ for $n\theta '\ll 1$, which
is pure thermal wandering. As $n$ increases further, the
potential starts to have an effect, and for $n\theta ' \simeq \pi/2$,
we find that the transverse scale increases linearly with
line length $\gamma _{n} \sim l n $, with a prefactor depending
in a non-trivial way on $\chi $.

\section{Conclusions}

In this paper we have studied in detail the action of a pulsed
harmonic potential on three systems: a classical particle,
a quantum particle, and a directed line. The first and second
systems share some properties via their mechanics, whereas the 
second and third share a common mathematical basis via the path
integral formalism. The pulsing was taken to be regular with a
period $\tau $ (or a longitudinal wavelength $d$ in the case
of a directed line).

The classical particle was studied in section II. It was
found to have stable (or bounded)
dynamics as long as the dimensionless coupling $\xi $ (cf.
Eq.(\ref{dimcoup})) lies in the range $0 \le \xi \le 4$. 
Otherwise the motion is unstable. For $\xi < 0$, the particle is
accelerated to $|x|=\infty$, whilst for $\xi >4$, the particle
`ping-pongs' with ever increasing amplitude away from the origin.
In the stable band, there is either periodic or quasi-periodic
motion. We introduced two classes of periodic motion: PMI($n$),
for motion where all physical quantities have period $n\tau $;
and PMII($n$), for motion where the energy has period $n\tau $.
We found that the condition for PMI($n$) motion was that
$\phi = M\pi /n$ where $\phi $ is a convenient parameter
defined as $\cos \phi = 1-\xi/2$, and $M$ is a positive {\it even}
integer satisfying $M\le [n/2]$. Thus the simplest PMI motion occurs
for $n=3$ and $M=2$ (corresponding to $\xi=3$). 
PMII($n$) motion was found to
occur for $\phi = M'\pi /n$ with $M'$ an integer (even or odd)
in the range $1 \le M' \le [n/2]$. The simplest PMII motion occurs
for $n=2$ and $M'=1$ (corresponding to $\xi=2$). These periodic motions are
independent of the initial position $x_{1}$ and initial momentum
$p_{1}$. We also found one special PMI(2) motion which requires
$\xi=4$ and a specially tuned initial condition $p_{1}=2mx_{1}/\tau $.

The dynamical properties of a quantum particle in a PHP were 
studied in sections III-V. The first two sections concentrated on 
a gaussian wave packet centered at the origin; whilst section V
pertained to the case of an off-centered gaussian wave packet,
which has non-zero expectation values of position and momentum,
and may therefore be compared directly to the classical particle
studied in section II.

In section III we studied the centered gaussian wave packet in
a PHP using the iteration properties of the determinants of
$n$-fold gaussian integrals. This led to second order linear
iteration rules similar to those found in section II. The
stability band for the wave packet is $0 \le \xi < 4$ as in
the classical case. If $\xi < 0$, the wave packet is stretched
more with each pulse and the width monotonically diverges with time.
If $\xi > 4$, the wave packet is squeezed so tightly after each
pulse that the velocity of expansion of the width of the packet
grows ever greater after each pulse (due to quantum uncertainty).
Within the stable band, there are periodic and quasi-periodic
dynamics. In the former, the cycles are periodic for all physical
properties, thus there is no classification into PMI and PMII
as in the classical case. The condition for a cycle of period
$n\tau $ is $\phi = M\pi /n$ with $M$ an integer (even or odd)
in the range $1 \le M \le [n/2]$, where $\phi $ is defined via
$\cos \phi = 1-\xi/2$ as used in the classical case. The energy
portraits for the classical and quantum particles are shown
in Figs. 1 and 2, and show several distinct features, especially
with regard to the upper and lower bounding curves.

In section IV we analysed the centered gaussian wave packet
using Fourier methods, which resulted in a first-order,
but nonlinear, iteration rule. We showed its equivalence
to the second-order linear iteration rule of section III.
This first order rule allowed two new aspects of the problem
to be analysed with ease. The first is the existence of a
special cycle of period $\tau $, which is a purely quantum
mechanical effect, as a classical system must have a cycle
of at least $2\tau $. This 1-cycle exists only for an initial
complex gaussian wave function with an inverse variance tuned to the
harmonic coupling via ${\bar \eta} = 2\sin (\phi /2)\exp (i\phi/2)$.
An example of a 1-cycle is shown in Fig. 5.
The second aspect is the behaviour of the system as the time $\tau $
between pulses is taken to zero. We found that if the initial wave 
function is chosen to be the ground state of a static harmonic
potential (with oscillator frequency $\omega = (\lambda /m\tau )^{1/2}$), 
then this wave function is a fixed point of the PHP dynamics as
$\tau \rightarrow 0$. This result shows that a pulsed potential
with arbitrarily small period cannot mimic a static potential, 
for even though the ground state wave function is a fixed point
of the PHP dynamics, there is no phase evolution as would be
found for a wave function in the static potential. The PHP continually
resets the phase clock. This causes concern with regard to numerical
integration of the time-dependent Schr\"odinger equation, where
one may discretize time, in which case one is implicitly modelling
a static potential by a pulsed potential.

In section V we allowed the gaussian wave packet to be off-centre,
which allowed there to be an evolution of the expectation values
${\bar x}$ and ${\bar p}$ of the position and momentum respectively. 
Using our previous iteration rules
in tandem, we showed that these expectation values obeyed the
same difference equations as the classical position and momentum,
as studied in section II, thus verifying Ehrenfest's theorem in
a non-trivial setting. Furthermore, we evaluated the energy of
the wave packet, and found that it split neatly into two pieces:
a `quantum piece' equal to the energy of the centered
gaussian wave packet, and a `classical piece' equal to
${\bar p}^{2}/2m$.

We moved away from the quantum PHP, and in section VI studied 
the statistical mechanics of a directed line in a harmonic
potential which exists only on discrete transverse lines which
are regularly spaced in the longitudinal direction. This system
is the imaginary time analogue of the quantum system, as is
clear from the path integral formalism. Using the same methods
as in section III, we found that this system has two qualitatively
different regimes, depending on the dimensionless coupling
${\tilde g}$ (which is the analogous quantity to $\xi $ as used in
the mechanical systems). For ${\tilde g}>0$, we found that the
transverse fluctuations of the line saturate rapidly to
$l/\sqrt{\zeta } $, where $l$ is the transverse thermal wandering scale
between pulses, and $\zeta = 1 - e^{-\theta }$, where $\cosh \theta
=1 + {\tilde g}$. For very large attractive coupling, the transverse
scale saturates at $l$ with $O(1/{\tilde g})$ corrections.
For ${\tilde g} < 0$ the line has a maximum length $n^{*}d$ beyond
which it is destroyed (as a connected elastic entity) by the
repulsive PHP. An exact expression was derived for $n^{*}$
(as given in Eq.(\ref{upperlen})), which has the asymptotic form
for $|{\tilde g}| \rightarrow 0$: $n^{*} = [w^{*}]$, with 
$w^{*} \sim \pi/|{\tilde g}|^{1/2}$. Prior to the line breaking
up, the transverse wandering grows diffusively for 
$n|{\tilde g}|^{1/2} \ll 1$, and linearly with $n$ for $
n|{\tilde g}|^{1/2} \simeq \pi/2$.

These results show clearly the subtle differences which exist between
the mechanics of the classical and quantum particles in a PHP; and
also the differences which exist between the statistics of
the quantum and statistical path integrals expressions for the PHP.
In the former case we have seen that the quantum system has another
level of complexity beyond the mean (or classical) motion.
The periodic motion of the centered gaussian
wave function has no classical counterpart, and its energy portrait
certainly deserves more study. In the latter case of the quantum 
versus statistical path integrals, we have seen how no vestige of
the periodic behaviour of the quantum system remains in the physics 
of the directed line. It has a much simpler asymptotic ($n \rightarrow
\infty$) behaviour, since all the interesting quantum effects are here
damped exponentially in $n$. This serves as a warning that one can
only retrieve complete quantum information from an imaginary time path 
integral, if one has complete analytic information.

We believe that these results may also be of some practical interest.
It is well known how to trap single particles in specially prepared
potentials formed from external magnetic fields\cite{trap}. 
Thus it is possible to make the trapping potential
time dependent by externally varying these fields. It
would be of interest to use such external variation to mimic a PHP,
and to test the generality of the results obtained here; namely,
the stability band, and its associated periodic and quasi-periodic
dynamics. 

As detailed in section VI, one can find applications
for the directed line in a PHP in the field of superconductivity.
Of more recent interest in this field is the role of disorder in
layered materials\cite{dis}, and its efficacy in pinning flux lines. Disorder
is a notoriously difficult effect to describe analytically, and there
are essentially no analytically solvable cases of directed lines in a quenched
disorder potential. We consider the generalization of the regularly
pulsed PHP to one with random pulsing intervals 
(along with a localized columnar pin) to be a prime candidate
for such a solvable system, using the framework developed in this paper.

\vspace{1.0cm}
The authors gratefully acknowledge financial support from the
Division of Materials Research of the National Science Foundation.

\newpage

\begin{figure}[htbp]
\centerline{\epsfxsize=20.0cm
\epsfbox{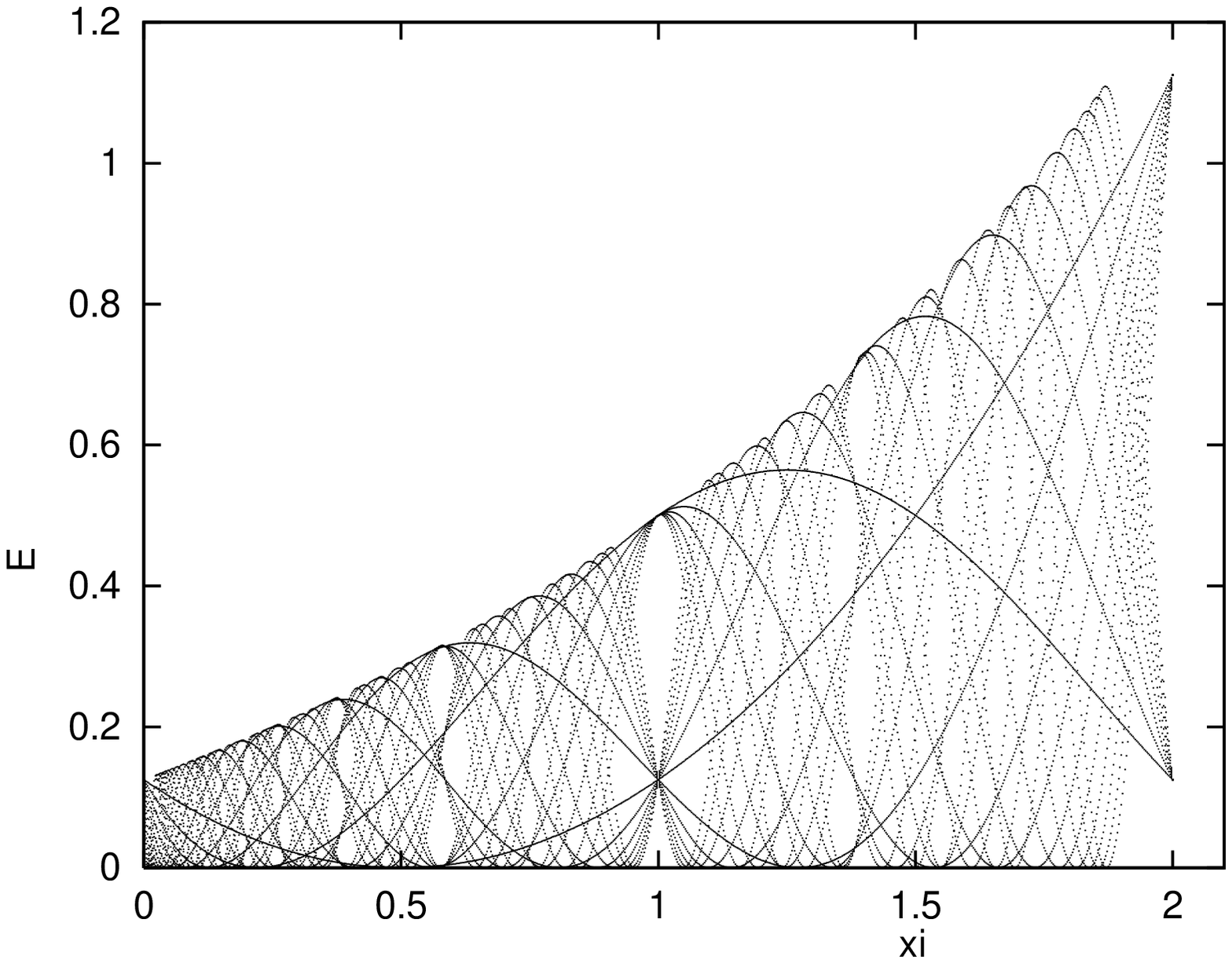}}
\vspace{0.4cm}
\caption{Energy portrait for a classical particle of unit mass in a PHP 
for $0 \le \xi \le 2$, and initial conditions $x(0)=1$ and $\rho _{1}=1/2$.}
\end{figure}
\vspace{2.cm}

\begin{figure}[htbp]
\centerline{\epsfxsize=10.0cm
\epsfbox{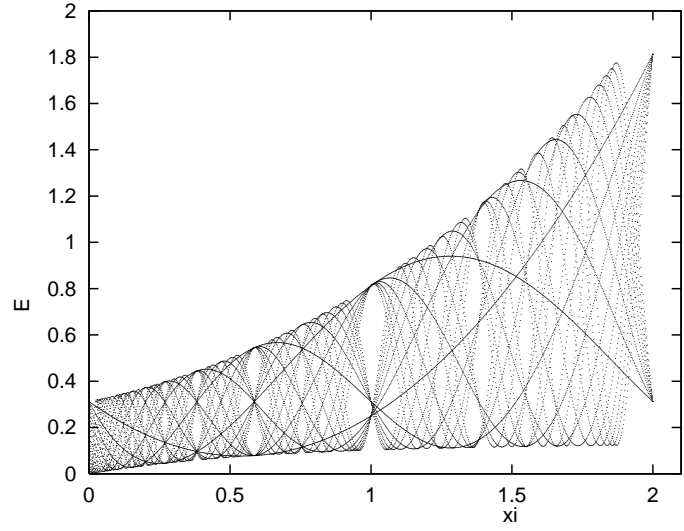}}
\vspace{0.4cm}
\caption{Energy portrait for a quantum particle in a PHP for
$0 \le \xi \le 2$ with $\hbar = \tau =1$, and $\eta = 1-i/2$.}
\end{figure}

\begin{figure}[tbp]
\centerline{\epsfxsize=10.0cm
\epsfbox{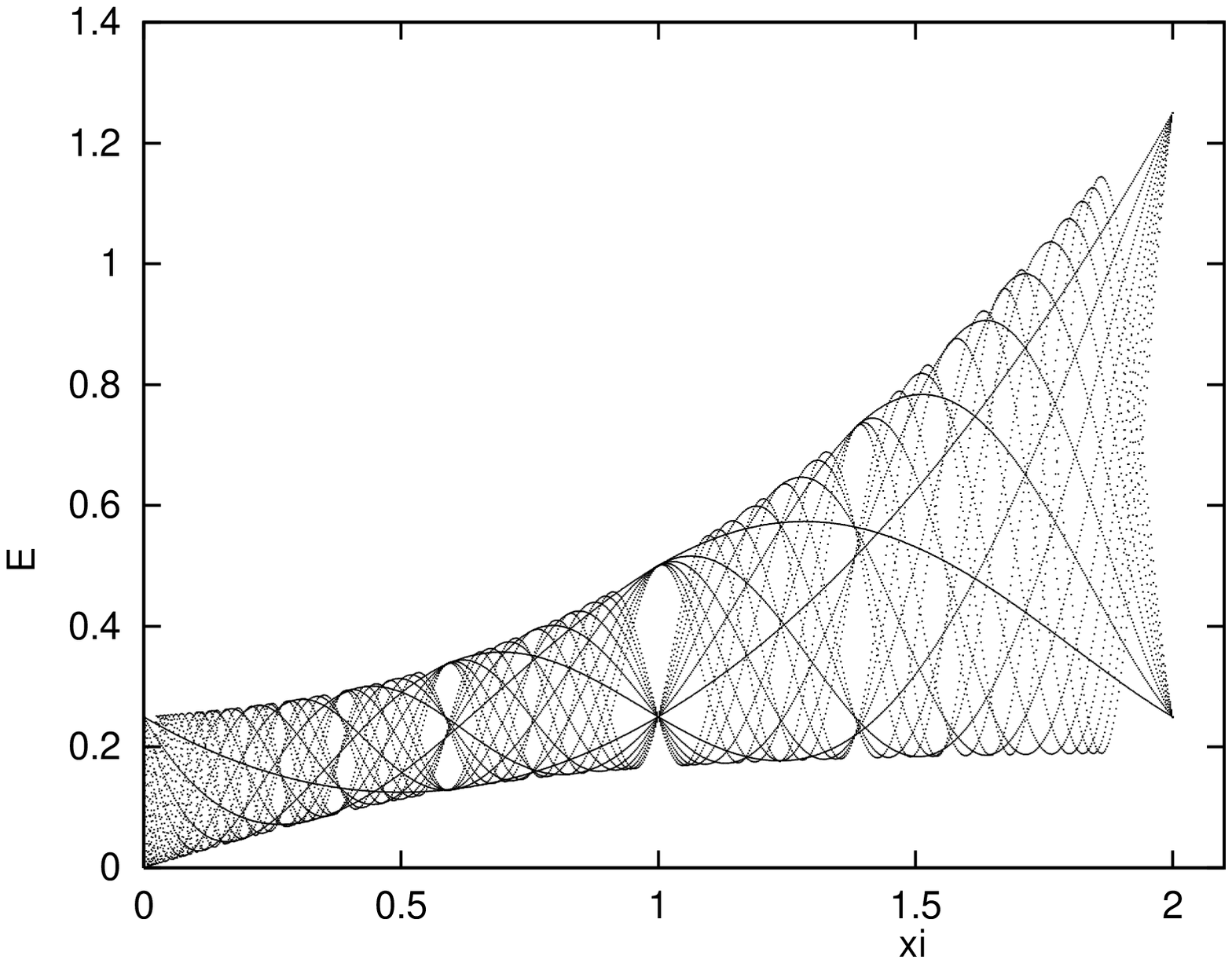}}
\vspace{0.4cm}
\caption{As Fig.2 with $\eta = 1$.}
\end{figure}

\begin{figure}[tbp]
\centerline{\epsfxsize=10.0cm
\epsfbox{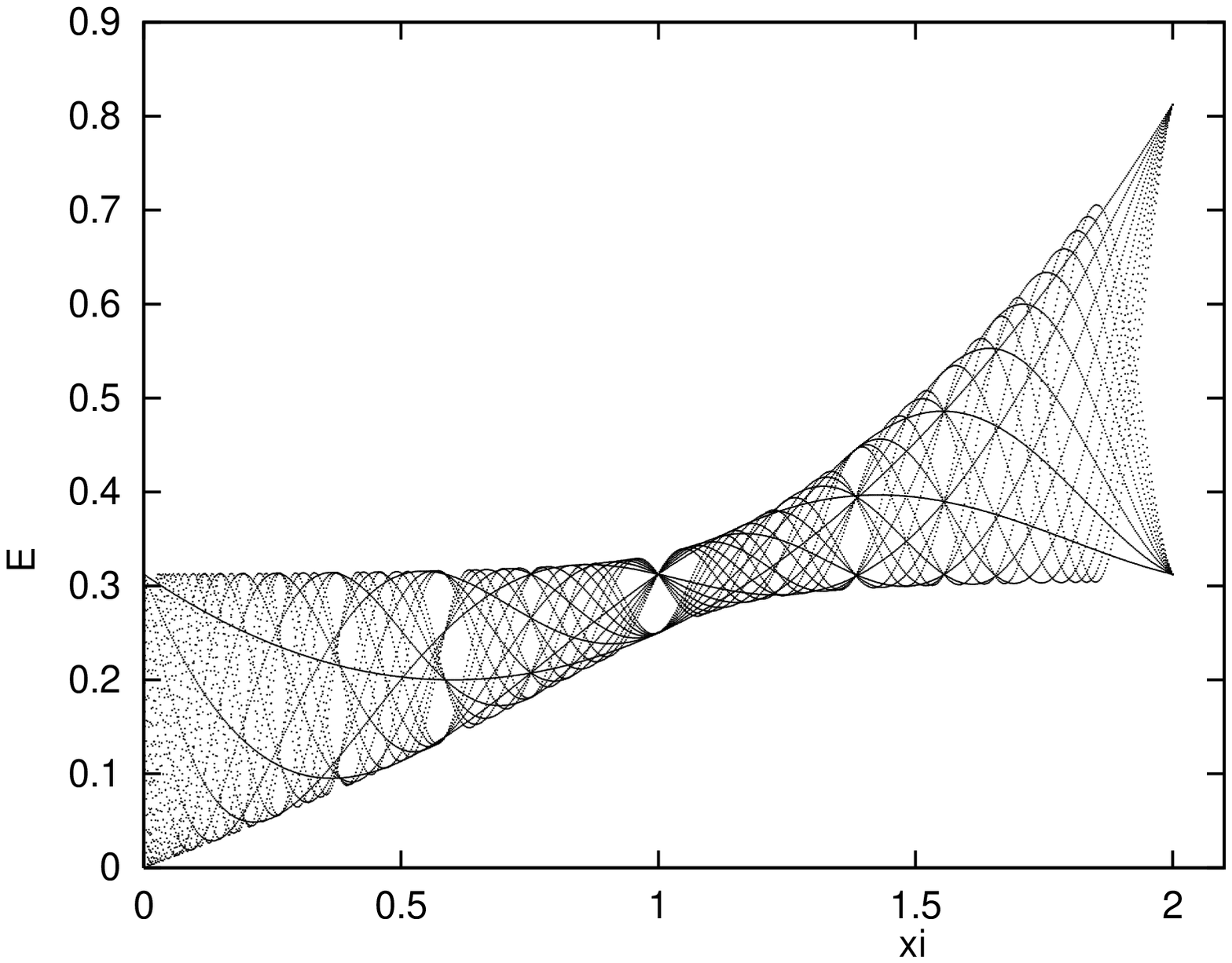}}
\vspace{0.4cm}
\caption{As Fig.2 with $\eta = 1+i/2$.}
\end{figure}

\begin{figure}[tbp]
\centerline{\epsfxsize=10.0cm
\epsfbox{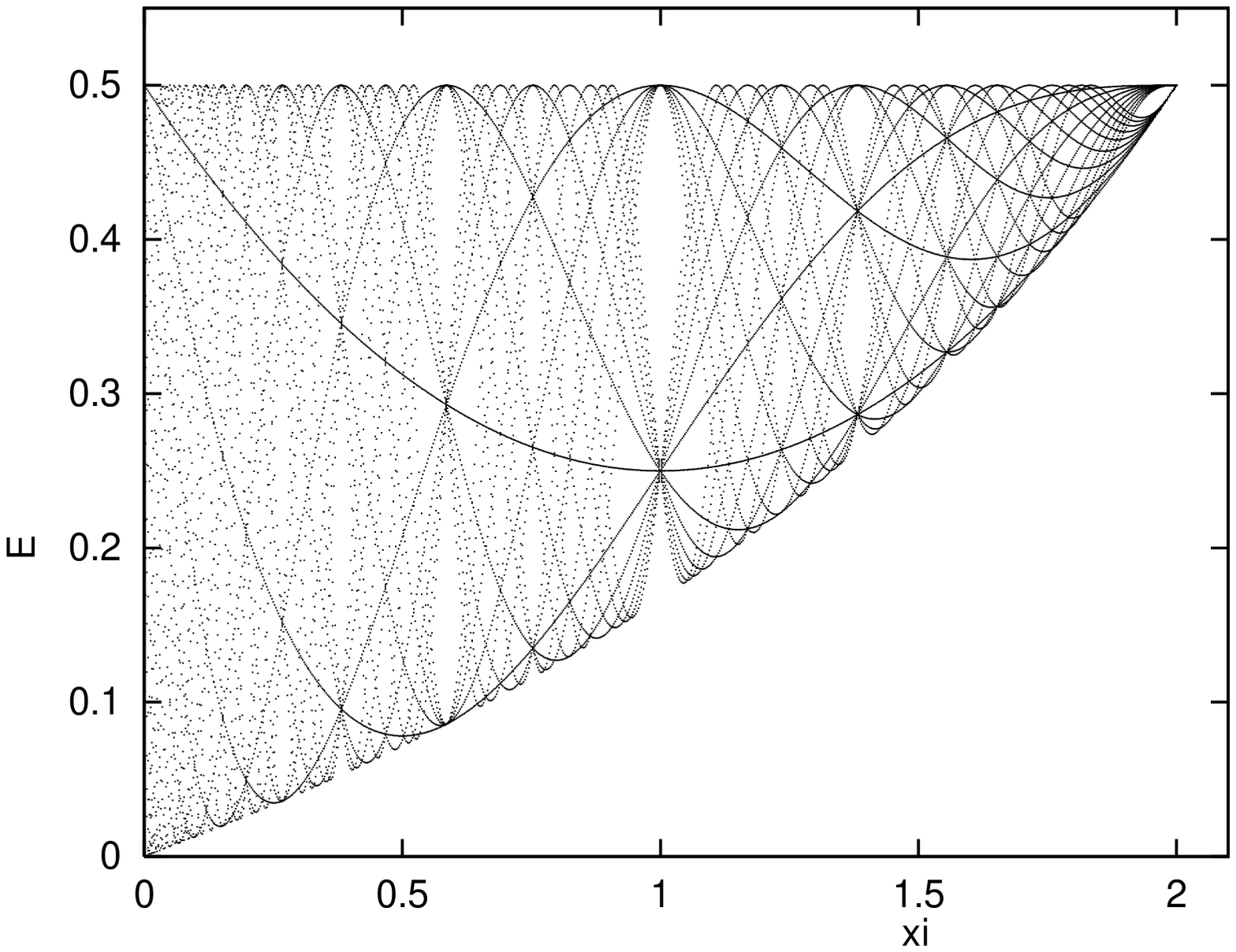}}
\vspace{0.4cm}
\caption{As Fig.2 with $\eta = 1+i$. Note the 1-cycle at $\xi =2 $.}
\end{figure}


\newpage

\begin{references}

\bibitem{ditt} W. Dittrich and M. Reuter, {\it Classical and Quantum 
Dynamics}, (Springer-Verlag, Berlin, 1992).

\bibitem{gard} N. G. van Kampen, {\it Stochastic Processes in Physics 
and Chemistry}, 2nd ed. (North Holland, Amsterdam, 1992).

\bibitem{class} V. I. Arnold, {\it Mathematical Methods of Classical
Mechanics} 2nd ed. (Springer-Verlag, New York, 1989).

\bibitem{sch} E. Schr\"odinger, Naturwissenschaften, {\bf 14}, 664 (1926).

\bibitem{har}J. G. Hartley and J. R. Ray, Phys. Rev. D, 
{\bf 25} 383 (1982).

\bibitem{hag} G. A. Hagedorn, M. Loss and J. Slawny, J. Phys. A, {\bf 19}
521 (1986).

\bibitem{sik} A. K. Sikri, S. C. Gupta and M. L. Narchal, Ind. J. Phys,
{\bf 67B} 15 (1993). 

\bibitem{gutz} M. C. Gutzwiller, {\it Chaos in Classical and Quantum 
Mechanics} (Springer-Verlag, New York, 1990).

\bibitem{berry} J. H. Hannay and M. V. Berry, Physics {\bf 1D} 267 (1980).

\bibitem{ford} J. Ford, in {\it Directions in Chaos}, ed. H. Bai-lin
(World Scientific, Singapore, 1988).

\bibitem{kho} G. P. Berman, V. Yu. Rubaev and G. M. Zaslavsky, 
Nonlinearity, {\bf 4} 543 (1991); S. A. Gardiner, J. I. Cirac and P. Zoller,
Phys. Rev. Lett., {\bf 79} 4790 (1997).

\bibitem{new} T. J. Newman and A. J. McKane, Phys. Rev. E, 
{\bf 55} 165 (1997).

\bibitem{pi} R. P. Feynman and A. Hibbs, {Quantum Mechanics and 
Path Integrals} (McGraw-Hill, New York, 1965); L. S. Schulman, {\it
Techniques and Applications of Path Integration} (Wiley, New York, 1981).

\bibitem{schiff} L. I. Schiff, {\it Quantum Mechanics} 3rd Edition 
(McGraw-Hill, Singapore, 1968).

\bibitem{vit} D. Vitali and P. Tombesi, preprint quant-phys/9808055 (1998).

\bibitem{dp} T. Halpin-Healy and Y.-C. Zhang, Phys. Rep. {\bf 254}, 
215 (1995).

\bibitem{blat} G. Blatter et. al., Rev. Mod. Phys. {\bf 66}, 1125 (1994). 

\bibitem{cup} J. C. Phillips, {\it Physics of High-Temperature
Superconductors}, (Academic Press, San Diego, 1989).

\bibitem{trap} See for example H. Dehmelt, Rev. Mod. Phys, {\bf 62} 525
(1990); W. Paul, {\it ibid}, {\bf 62}, 531 (1990).

\bibitem{dis} {\it Phenomenology and Applications of High-Temperature
Superconductors}, eds. K. Bedell et al., (Addison-Wesley, New York, 1992).

\end{references}
\end{document}